\documentclass[11pt]{article}
\usepackage{amssymb,amsmath}
\usepackage{graphicx}
\usepackage{epsfig}

\input{tcilatex}

\begin{document}

\title{Structure of Noncommutative Fock space
\thanks{%
This project supported by National Natural Science Foundation of
China under Grant 10375056 and 90203002.}}
\author{{Sicong Jing, Qiuyu Liu and Tunan Ruan} \\
%EndAName
{\small Department of Modern Physics, University of Science and Technology
of China,}\\
{\small Hefei, Anhui 230026, China}} \maketitle

\begin{abstract}
The structure of the state-vector space of identical bosons in
noncommutative spaces is investigated. To maintain Bose-Einstein
statistics the commutation relations of phase space variables
should simultaneously include coordinate-coordinate
non-commutativity and momentum-momentum non-commutativity, which
lead to noncommutative Fock space. By this we mean that creation
and annihilation operators corresponding to different degrees of
freedom of the bosons do not commute each other. The main
character of the noncommutative Fock space is there are no
ordinary number representations because of the non-commutativity
between different number operators. However, eigenvectors of
several pairs of commuting Hermitian operators are obtained which
can also be served as bases in this Fock space. As a simple
example, an explicit form of two-dimensional canonical coherent
state in this noncommutative Fock space is constructed and its
properties are discussed.
\newline

\end{abstract}

\section{Introduction}
Recently there has been a renewed interest in quantum field theory
in noncommutative spaces (NCQFT)\cite{s1}-\cite{s4}, mainly due to
their relation to string theory in nontrivial backgrounds
\cite{s5} and M-theory compactifications \cite{s6}. It is usually
believed that the effects of noncommutative spaces only appear
near the string scale, thus to test the effects of space
non-commutativity NCQFT is needed. As in the commutative case, the
low energy limit of NCQFT few-particle sector can be taken which
results in noncommutative counterpart of ordinary quantum
mechanics. In literature the perturbation aspects of
noncommutative quantum mechanics (NCQM) have been extensively
studied \cite{s7}-\cite{s12}. The perturbation approach is based
on the Weyl-Moyal correspondence \cite{s13}, according to which
the usual product of functions on the noncommutative space should
be replaced by star-product. In the non-perturbation aspects of
NCQM, on the other hand, several essentially new features have
been explored by some authors \cite{s14}.

\par
A suitable example for NCQM is two-dimensional isotropic harmonic
oscillator, which is exactly solvable and fully explored
\cite{s15,s16}. The state-vector space of identical bosons in the
noncommutative space can be constructed by generalizing ordinary
quantum mechanics. In order to maintain Bose-Einstein statistics
for the identical bosons at the non-perturbation level one should
consider both coordinate-coordinate non-commutativity and
momentum-momentum non-commutativity, which result in a new type of
boson commutation relations \cite{s14}. The main difference
between the new type of boson algebraic relations and the ordinary
ones lies in that creation and annihilation operators
corresponding to different degrees of freedom of the bosons do not
commute each other, thus the different number operators are
non-commuting and have no common eigenvectors. In this sense, we
may name the relevant state-vector space as noncommutative Fock
space.

\par
In the present letter we analyze the structure of the
noncommutative Fock space. Although the different number operators
are noncommutative and have no simultaneously eigenvectors, we can
still construct the state-vectors as usual. It is worth pointing
out here, however, that the set of the such constructed
state-vectors is not suitable to be used as a set of basis vectors
for the noncommutative Fock space, because it is neither
orthogonal nor complete. In order to obtain complete set of
eigenvectors of some Hermitian operators which can be served as
the bases of the noncommutative Fock space, we investigate several
pairs of commuting Hermitian operators and find out their
simultaneously eigenvectors respectively. As a simple example, we
also construct two-dimensional canonical (coordinate-momentum)
coherent states in this noncommutative Fock space and study their
properties.

\par
The work is arranged as follows. In Section 2 we briefly review
that in order to maintain Bose-Einstein statistics for identical
particles on a noncommutative space at the level of
creation-annihilation operators, we should consider both
coordinate-coordinate non-commutativity and momentum-momentum
non-commutativity, which lead to a new type of algebraic relations
of creation and annihilation operators. The state-vector space can
be constructed by generalizing the ordinary quantum mechanics.
Since these state-vectors are not orthogonal each other, we use
the Schmidt's orthogonalization method to obtain sets of
orthonormal state-vectors in Section 3, which can be served as
bases in the noncommutative Fock space. We also consider several
pairs of commuting Hermitian operators and find out their
simultaneously eigenvectors in the noncommutative space. In
Section 4, as a simple application, we construct two-dimensional
coherent states in the noncommutative space and prove their
over-completeness property. We also show that the Heisenberg
coordinate-momentum uncertainty relations in NCQM are saturated by
the constructed coherent states and the other two uncertainty
relations (coordinate-coordinate and momentum-momentum) are not.
Some summary and discussion are in Section 5.

\section{Identical bosons in noncommutative space}
We start from the following consistent NCQM algebraic relations
(henceforth, we take $\hbar$ to be unit) \cite{s14}
\begin{eqnarray}
&&[\hat{x},\,\hat{y}]=\emph{i}\,\xi^{-2}\Lambda^{-2}d,~~~~
[\hat{p}_{x},\,\hat{p}_{y}]=\emph{i}\,\xi^{-2}\Lambda^{2}d',
\nonumber\\
&&[\hat{x},\,\hat{p}_{x}]=\emph{i},~~~~
[\hat{y},\,\hat{p}_{y}]=\emph{i},~~~~
[\hat{x},\,\hat{p}_{y}]=0,~~~~ [\hat{y},\,\hat{p}_{x}]=0,
\end{eqnarray}
where $d$ and $d'$ are frame-independent dimensionless constants,
$\Lambda$ is the NC energy scale \cite{s18} (usually
$\theta=\Lambda^{-2}d$ is adopted in literature for the case of
only coordinate-coordinate non-commuting). In eq.(1) $\xi$ is a
scaling factor $\xi=(1+d d'/4)^{1/2}$. When $d'=0$, eq.(1) reduces
to the algebraic relation which is extensively discussed in
literature for the case that only coordinate-coordinate are
non-commuting.

\par
The NCQM algebra (1) changes the usual algebra of bosonic
creation-annihilation operators. There are different ways to
construct the creation-annihilation operators for the algebra (1).
In order to explore noncommutative effects at the non-perturbation
level we introduce the creation-annihilation operators which
relate to the noncommutative phase space variables as follows
\begin{eqnarray}
&&\hat{a}=\sqrt{\frac{m\omega}{2}}\,\left(\hat{x}+\frac{\emph{i}}{m\omega}
\hat{p}_{x} \right),~~~~
\hat{a}^{\dag}=\sqrt{\frac{m\omega}{2}}\,\left(\hat{x}-\frac{\emph{i}}{m\omega}
\hat{p}_{x} \right), \nonumber\\
&&\hat{b}=\sqrt{\frac{m\omega}{2}}\,\left(\hat{y}+\frac{\emph{i}}{m\omega}
\hat{p}_{y} \right),~~~~
\hat{b}^{\dag}=\sqrt{\frac{m\omega}{2}}\,\left(\hat{y}-\frac{\emph{i}}{m\omega}
\hat{p}_{y} \right).
\end{eqnarray}
The state-vector space of identical bosons of NCQM can be
constructed by generalizing ordinary commuting quantum mechanics.
In order to maintain Bose-Einstein statistics at the
non-perturbation level described by $\hat{a}^{\dag}$ and
$\hat{b}^{\dag}$, a consistency condition should be satisfied
\cite{s14}
\begin{equation}
d'=m^{2}\omega^{2}\Lambda^{-4}d.
\end{equation}
Therefore eqs.(1-3) lead to the following commutation relations
for the creation and annihilation operators
\begin{equation}
[\hat{a},\,\hat{a}^{\dag}]=1,~~~~[\hat{b},\,\hat{b}^{\dag}]=1,~~~~
[\hat{a},\,\hat{b}]=0,~~~~[\hat{a}^{\dag},\,\hat{b}^{\dag}]=0,
\end{equation}
and
\begin{equation}
[\hat{a},\,\hat{b}^{\dag}]=\emph{i}\,\xi^{-2} m\omega
\Lambda^{-2}d.
\end{equation}
Henceforth we still use the notation $\theta$ (which is a real
number) to indicate $\xi^{-2} m\omega \Lambda^{-2}d$ and rewrite
eq.(5) as
\begin{equation}
[\hat{a},\,\hat{b}^{\dag}]=\emph{i}\,\theta, ~~~~(\theta=\xi^{-2}
m\omega \Lambda^{-2}d).
\end{equation}
Thus we can treat the algebraic relations (4) and (6) as a kind of
deformation of usual bosons commutation relations and regard
$\theta$ as the deformation parameter. From eq.(4) we know that
for the same degree of freedom of phase space variables (for
example, $\hat{x}$ and $\hat{p}_{x}$, or $\hat{y}$ and
$\hat{p}_{y}$), the relevant creation and annihilation operators
satisfy the same commutation relations as the ones in the
commutative space and this is very reasonable. The new commutation
relation (6) (or (5)) of creation and annihilation operators only
emerges for the different degree of freedom of bosons and
represents the effects of non-commutativity. Therefore we can
define number operators for each degree of freedom of bosons as
usual
\begin{equation}
\hat{N}_{a}=\hat{a}^{\dag}\,\hat{a},~~~~~~~~
\hat{N}_{b}=\hat{b}^{\dag}\,\hat{b}.
\end{equation}
It is easily to see that $[\hat{N}_{a},\,\hat{N}_{b}] \not= 0$, so
there exist no simultaneously eigenvectors of $\hat{N}_{a}$ and
$\hat{N}_{b}$.

\par
In order to construct the state-vector space for the above bosonic
system in noncommutative space, we need a unique normalized vacuum
state $|00>$ which is defined as $\hat{a}\,|00>=0$ and
$\hat{b}\,|00>=0$. Since $[\hat{a}^{\dag},\,\hat{b}^{\dag}]=0$, a
state with $n$ $\hat{a}$ bosons and $m$ $\hat{b}$ bosons can be
construct as in the case of commutative space,
\begin{equation}
|n,m>=N_{n,m}(\theta)\,\hat{a}^{\dag n}\,\hat{b}^{\dag m}\,|00>,
\end{equation}
where $N_{n,m}(\theta)$ is a normalization factor. Therefore, the
set of states $\{|n,m>|n,m=0,1,2,,...\}$ can span the state-vector
space. In contrast to the case of commutative space, however, here
the set $\{|n,m>|n,m=0,1,2,,...\}$ can not be used as a suitable
set of basis vectors for this noncommutative Fock space, because
the state-vector $|n,m>$ are neither orthogonal each other nor
complete (for the detail, see the next section). By the latter we
mean that we have not the relation $\sum_{n,m=0}^{\infty}
\,|n,m>\,<n,m|=1$. Generally speaking, the normalization factor
$N_{n,m}(\theta)$ in $|n,m>$ is a function of the deformation
parameter $\theta$. As mentioned in Section 1, because usually the
effects of noncommutative space only appear near the string scale,
so one can take the deformation parameter $\theta$ as a very small
quantity. Up to the first power of $\theta$, it is easily to get
$N_{n,m}(\theta)=1/\sqrt{n!\,m!}$, so the normalized state-vector
$|n,m>$ has the familiar form
\begin{equation}
|n,m>=\frac{1}{\sqrt{n!\,m!}}\,\hat{a}^{\dag n}\,\hat{b}^{\dag
m}\,|00>.
\end{equation}
When the creation and annihilation operators $\hat{a}^{\dag}$,
$\hat{b}^{\dag}$, $\hat{a}$ and $\hat{b}$ act on the state
$|n,m>$, one has
\begin{eqnarray}
&&\hat{a}\,|n,m>=\sqrt{n}\,|n-1,m>+\emph{i}\,\theta\,\sqrt{m}\,|n,m-1>,
~~~~\hat{a}^{\dag}\,|n,m>=\sqrt{n+1}\,|n+1,m>, \nonumber\\
&&\hat{b}\,|n,m>=\sqrt{m}\,|n,m-1>-\emph{i}\,\theta\,\sqrt{n}\,|n-1,m>,
~~~~\hat{b}^{\dag}\,|n,m>=\sqrt{m+1}\,|n,m+1>.
\end{eqnarray}
Here we find that the the annihilation operator $\hat{a}$ acting
on the state $|n,m>$ can destroy not only a boson $\hat{a}$, but
also a boson $\hat{b}$, and the same thing will occur for the
annihilation operator $\hat{b}$. Thus the state $|n,m>$ is neither
an eigenvector of the operator $\hat{N}_{a}$ nor one of
$\hat{N}_{b}$. As a result of this fact, the state-vectors in the
set of $\{|n,m>|n,m=0,1,2,...\}$ will be generally not orthogonal
each other. By virtue of the well-known Schmidt's
orthogonalization method, one can get sets of orthonormal
state-vectors which can be used as bases for the noncommutative
Fock space.

\section{The sets of basis vectors in the Noncommutative Fock space}
Consider a subspace of the noncommutative Fock space with total
$N$ identical bosons $\hat{a}$ and $\hat{b}$, and denote it as
$\emph{H}_{N}$. In a state $|n,m>$ belonging to $\emph{H}_{N}$
there are $n$ $\hat{a}$ bosons and $m$ $\hat{b}$ bosons, with
$0\leq n,\,m \leq N$ and $n+m=N$. We denote each partition of $N$
by $(n,m)$. For a given $N$, the number of its partitions is
$N+1$, which means that there are $N+1$ different states in
$\emph{H}_{N}$, i.e. the states $\{|n,m>|(n,m); 0 \leq n,\,m \leq
N, n+m=N \}$. For arbitrary two states belonging to different such
subspaces, the orthogonality is obvious: $<n,m|n',m'> \,\propto
\delta_{n+m,n'+m'}$. Two different states within a same such
subspace, however, are generally not orthogonal each other. After
simply analysis we find that up to the first power of $\theta$,
only for two nearest neighboring states their scalar products are
not vanishing
\begin{equation}
<n,m|n-1,m+1>=\emph{i}\,\theta \sqrt{n(m+1)}.
\end{equation}
By virtue of the Schmidt's orthogonalization method, we can
combine and rearrange the $N+1$ states in the subspace
$\emph{H}_{N}$ and get the following $N+1$ orthonormal states
\begin{equation}
|n;n+m>_{a}=|n,m>-\emph{i}\,\theta \sqrt{m(n+1)}\,|n+1,m-1>,~~~~~~
(0\leq n,\,m \leq N, ~~~n+m=N)
\end{equation}
which up to the first power of $\theta$ satisfy
\begin{equation}
\hat{N}_{a}|n;n+m>_{a}=n|n;n+m>_{a},
\end{equation}
and are the eigenvectors of the number operator $\hat{N}_{a}$.
Similarly, one can get another $N+1$ orthonormal states in
$\emph{H}_{N}$
\begin{equation}
|m;n+m>_{b}=|n,m>+\emph{i}\,\theta \sqrt{n(m+1)}\,|n-1,m+1>,~~~~~~
(0\leq n,\,m \leq N, ~~~n+m=N)
\end{equation}
and which up to the first power of $\theta$ satisfy
\begin{equation}
\hat{N}_{b}|m;n+m>_{b}=m|m;n+m>_{b}.
\end{equation}
Thus the state-vectors $|n;n+m>_{a}$ (or $|m;n+m>_{b}$) are
eigenvectors of the number operators $\hat{N}_{a}$ (or
$\hat{N}_{b}$) with $n$ $\hat{a}$ (or $m$ $\hat{b}$) bosons
respectively, and form complete and orthonormal sets of
eigenvectors .

\par
Obviously, the eigenvectors $\{|n;n+m>_{a}|n,m=0,1,2,... \}$ (or
$\{|m;n+m>_{b}|n,m=0,1,2,... \}$) of the number operator
$\hat{N}_{a}$ (or $\hat{N}_{b}$) are infinite degenerate for the
eigenvalues $n$ (or $m$). In order to remove these degeneracies
let us introduce a Hermitian operator
\begin{equation}
\hat{L}=\emph{i}\,(\hat{a}^{\dag}\hat{b}-\hat{b}^{\dag}\hat{a}),
\end{equation}
which is usually called the angular momentum operator. From the
starting algebraic relations (4) and (6), one can easily find that
the operator $\hat{N}_{a}+\hat{N}_{b}-\theta \hat{L}$ commutes
with both $\hat{N}_{a}$ and $\hat{N}_{b}$, so using this operator
one can eliminate the above degeneracies. In fact, the
state-vectors $|n;n+m>_{a}$ (or $|n;n+m>_{b}$) are the
simultaneously eigenvectors of $\hat{N}_{a}$ (or $\hat{N}_{b}$)
and $\hat{N}_{a}+\hat{N}_{b}-\theta \hat{L}$ with eigenvalues $n$
(or $m$) and $n+m$ respectively (up to the first power of
$\theta$)
\begin{eqnarray}
\hat{N}_{a}|n;n+m>_{a}=n|n;n+m>_{a},
&&(\hat{N}_{a}+\hat{N}_{b}-\theta
\hat{L})|n;n+m>_{a}=(n+m)|n;n+m>_{a}, \nonumber\\
\hat{N}_{b}|m;n+m>_{b}=m|m;n+m>_{b},
&&(\hat{N}_{a}+\hat{N}_{b}-\theta
\hat{L})|m;n+m>_{b}=(n+m)|m;n+m>_{b}.
\end{eqnarray}
Thus we have two sets of orthonormal state-vectors
$\{|n;n+m>_{a}\}$ and $\{|m;n+m>_{b}\}$, which are eigenvectors of
commuting Hermitian operator pairs of
$\hat{N}_{a}+\hat{N}_{b}-\theta \hat{L}$ and $\hat{N}_{a}$ or
$\hat{N}_{b}$, so can be used as the bases for the noncommutative
Fock space because of their completeness properties
\begin{equation}
\sum_{n,m=0}^{\infty}\,|n;n+m>_{a\,a}<n;n+m|=1, ~~~~
\sum_{n,m=0}^{\infty}\,|m;n+m>_{b\,b}<m;n+m|=1.
\end{equation}
Substituting eq.(12) into the first expression of eq.(18), also up
to the first power of $\theta$, we have
\begin{equation}
\sum_{n,m=0}^{\infty}|n,m>\,<n,m|=1+\emph{i}\,\theta
\sum_{n,m=0}^{\infty}\sqrt{m(n+1)}\,\left(
|n+1,m-1>\,<n,m|-|n,m>\,<n+1,m-1| \right),
\end{equation}
which shows that the set of state-vectors ${|n,m>}$ is incomplete.

\par
Besides the commuting operator pairs of
$\hat{N}_{a}+\hat{N}_{b}-\theta \hat{L}$ and $\hat{N}_{a}$ or
$\hat{N}_{b}$, one can find other commuting operator pair also,
for example, $\hat{N}_{a}-\frac{\theta}{2}\hat{L}$ commutes with
$\hat{N}_{b}-\frac{\theta}{2}\hat{L}$. Their simultaneously
eigenvectors (up to the first power of $\theta$) are
\begin{equation}
|n,m>_{a,b}=|n,m>+\frac{\emph{i}\,\theta}{2}\sqrt{n(m+1)}\,|n-1,m+1>
- \frac{\emph{i}\,\theta}{2}\sqrt{m(n+1)}\,|n+1,m-1>
\end{equation}
which satisfy
\begin{equation}
\left(\hat{N}_{a}-\frac{\theta}{2}\hat{L}
\right)|n,m>_{a,b}=n|n,m>_{a,b},~~~~
\left(\hat{N}_{b}-\frac{\theta}{2}\hat{L}
\right)|n,m>_{a,b}=m|n,m>_{a,b}.
\end{equation}
Therefore the set of eigenvectors $\{|n,m>_{a,b}\}$ can also be
used as a set of basis vectors for this noncommutative Fock space.

\section{Two-dimensional coherent states}
Coordinate-coordinate coherent states in non-commutative space
have been discussed by some authors (for example, see
\cite{s14,s17}), however, two-dimensional canonical
(coordinate-momentum) coherent states in the noncommutative Fock
space have not reported in literature so far. In the
noncommutative Fock space, since $[\hat{a},\,\hat{b}]=0$, one can
construct their simultaneously eigenvectors, i.e. two-dimensional
canonical (coordinate-momentum) coherent states. Usually the
canonical coherent state has three equivalent definitions, as the
eigenvector of annihilation operator, as the minimal uncertainty
state of the Heisenberg uncertainty relation, or as the state
generated by a displacement operator acting on the vacuum state.
Here we define the two-dimensional canonical coherent state in
noncommutative Fock space by the third way and write it as
\begin{equation}
|\alpha,\beta>=N_{\alpha,\beta}(\theta)e^{\alpha\,\hat{a}^{\dag}}
e^{\beta\,\hat{b}^{\dag}} |00>,
\end{equation}
where $\alpha$ and $\beta$ are complex numbers, and
$N_{\alpha,\beta}(\theta)$ is a normalization factor. By virtue of
the following relations
\begin{equation}
e^{\beta^{\ast} \hat{b}}\hat{a}^{\dag}e^{-\beta^{\ast} \hat{b}}
=\hat{a}^{\dag}-\emph{i}\,\theta \beta^{\ast},~~~~
e^{\alpha^{\ast} \hat{a}}\hat{b}^{\dag}e^{-\alpha^{\ast} \hat{a}}
=\hat{b}^{\dag}+\emph{i}\,\theta \alpha^{\ast},
\end{equation}
one can find that
\begin{equation}
N_{\alpha,\beta}(\theta)=\exp{\left( \frac{1}{2}
\left(-|\alpha|^{2} -|\beta|^{2}-\emph{i}\,\theta \alpha^{\ast}
\beta +\emph{i}\,\theta \alpha \beta^{\ast} \right) \right)}
\end{equation}
and verify that $|\alpha,\beta>$ is indeed the simultaneously
eigenvector of $\hat{a}$ and $\hat{b}$ with the eigenvalue
$\alpha+\emph{i}\,\theta \beta$ and $\beta -\emph{i}\,\theta
\alpha$ respectively
\begin{equation}
\hat{a}|\alpha,\beta>=(\alpha+\emph{i}\,\theta
\beta)|\alpha,\beta>,~~~~ \hat{b}|\alpha,\beta>=(\beta
-\emph{i}\,\theta \alpha)|\alpha,\beta>.
\end{equation}
Furthermore, the scalar product of two such coherent states is
\begin{eqnarray}
<\alpha,\beta|\alpha',\beta'>&=&\exp{\left(
-\frac{1}{2}\left(|\alpha|^{2}
+|\beta|^{2}+|\alpha'|^{2}+|\beta'|^{2}\right)+\alpha^{\ast}\alpha'
+\beta^{\ast}\beta'\right)}   \nonumber\\
&&\exp{\left(\frac{\emph{i}\,\theta}{2}\left(\alpha
\beta^{\ast}-\alpha^{\ast}\beta +\alpha'
\beta'^{\ast}-\alpha'^{\ast} \beta'\right)+\emph{i}\,\theta
(\alpha^{\ast}\beta'-\alpha' \beta^{\ast})\right)},
\end{eqnarray}
which means that these coherent states are not orthogonal each
other, however, similarly to the case in commutative space, they
are also over-complete. In order to see this, using eq.(9), one
has
\begin{eqnarray}
\int\,\frac{d^{2} \alpha d^{2} \beta}{\pi^{2}}|\alpha,\beta>\,
<\alpha,\beta| = &&\int\,\frac{d^{2} \alpha d^{2} \beta}{\pi^{2}}
\exp{\left(-|\alpha|^{2} -|\beta|^{2} +\emph{i}\,\theta (\alpha
\beta^{\ast}-\alpha^{\ast} \beta)\right)} \nonumber\\
&&\sum_{n,m,n',m'=0}^{\infty} \frac{\alpha^{n}
\beta^{m}}{\sqrt{n!\,m!}}|n,m>\,<n,m|\frac{\alpha^{\ast
n'}\beta^{\ast m'}}{\sqrt{n'!\,m'!}}.
\end{eqnarray}
After integrating and up to the first power of $\theta$, one gets
\begin{equation}
\sum_{n,m=0}^{\infty}|n,m>\,<n,m| -\emph{i}\,\theta
\sum_{n,m=0}^{\infty}
\sqrt{m(n+1)}\,\left(|n+1,m-1>\,<n,m|-|n,m>\,<n+1,m-1|\right)=1,
\end{equation}
where eq.(19) is used. Here we should point out that since eqs.(9)
and (19) are only accurate up to the first power of $\theta$, the
calculation in eqs.(27) and (28) is also accurate up to this
order.

\par
Next we consider the uncertainty relations of NCQM for the above
introduced coherent states. The expectation values of the
operators $\hat{x}$, $\hat{y}$, $\hat{p}_{x}$ and $\hat{p}_{y}$ in
the state $|\alpha,\beta>$ are
\begin{eqnarray}
<\hat{x}>=\frac{1}{\sqrt{2 m \omega}}\left(\alpha+\alpha^{\ast}+
\emph{i}\,\theta (\beta -\beta^{\ast})\right), &&
<\hat{y}>=\frac{1}{\sqrt{2 m \omega}}\left(\beta+\beta^{\ast}-
\emph{i}\,\theta (\alpha -\alpha^{\ast})\right), \nonumber\\
<\hat{p}_{x}>=\sqrt{\frac{m \omega}{2}}\left(-\emph{i}(\alpha
-\alpha^{\ast})+\theta(\beta +\beta^{\ast}) \right), &&
<\hat{p}_{y}>=\sqrt{\frac{m \omega}{2}}\left(-\emph{i}(\beta
-\beta^{\ast})-\theta(\alpha +\alpha^{\ast}) \right),
\end{eqnarray}
and furthermore
\begin{eqnarray}
<\hat{x}^{2}>=\frac{1}{2m \omega}\left(1+(\alpha+\alpha^{\ast}
+\emph{i}\,\theta (\beta-\beta^{\ast}))^{2} \right),&&
<\hat{y}^{2}>=\frac{1}{2m \omega}\left(1+(\beta+\beta^{\ast}
-\emph{i}\,\theta (\alpha -\alpha^{\ast}))^{2}\right), \nonumber\\
<\hat{p}_{x}^{2}>=\frac{m \omega}{2}\left(1-(\alpha-\alpha^{\ast}+
\emph{i}\,\theta (\beta+\beta^{\ast}))^{2} \right),&&
<\hat{p}_{y}^{2}>=\frac{m
\omega}{2}\left(1-(\beta-\beta^{\ast}-\emph{i}\,\theta
(\alpha+\alpha^{\ast}))^{2} \right).
\end{eqnarray}
Thus the variances of the operators $\hat{x}$, $\hat{y}$,
$\hat{p}_{x}$ and $\hat{p}_{y}$ in the state $|\alpha,\beta>$ can
be obtained from
\begin{eqnarray}
(\triangle \hat{x})^{2}\equiv
<\hat{x}^{2}>-<\hat{x}>^{2}=\frac{1}{2m \omega},&& (\triangle
\hat{y})^{2}\equiv <\hat{y}^{2}>-<\hat{y}>^{2}=\frac{1}{2m
\omega}, \nonumber\\
(\triangle \hat{p}_{x})^{2}\equiv
<\hat{p}_{x}^{2}>-<\hat{p}_{x}>^{2}=\frac{m \omega}{2},&&
(\triangle \hat{p}_{y})^{2}\equiv
<\hat{p}_{y}^{2}>-<\hat{p}_{y}>^{2}=\frac{m \omega}{2},
\end{eqnarray}
which lead to
\begin{equation}
\triangle \hat{x}\,\triangle \hat{p}_{x}=\frac{1}{2},~~~~
\triangle \hat{y}\,\triangle \hat{p}_{y}=\frac{1}{2},~~~~
\triangle \hat{x}\,\triangle \hat{y}=\frac{1}{2m \omega},~~~~
\triangle \hat{p}_{x}\,\triangle \hat{p}_{y}=\frac{m \omega}{2}.
\end{equation}
Using the notation $\theta$ (in eq.(6)), one can rewrite eq.(1) as
\begin{eqnarray}
&&[\hat{x},\,\hat{y}]=\emph{i}\,\frac{\theta}{m\omega},~~~~
[\hat{p}_{x},\,\hat{p}_{y}]=\emph{i}\,m\omega \theta,
\nonumber\\
&&[\hat{x},\,\hat{p}_{x}]=\emph{i},~~~~
[\hat{y},\,\hat{p}_{y}]=\emph{i},~~~~
[\hat{x},\,\hat{p}_{y}]=0,~~~~ [\hat{y},\,\hat{p}_{x}]=0,
\end{eqnarray}
which imply the following uncertainty relations
\begin{equation}
\triangle \hat{x}\,\triangle \hat{p}_{x}\geq \frac{1}{2},~~~~
\triangle \hat{y}\,\triangle \hat{p}_{y}\geq \frac{1}{2},~~~~
\triangle \hat{x}\,\triangle \hat{y} \geq \frac{\theta}{2m
\omega},~~~~ \triangle \hat{p}_{x}\,\triangle \hat{p}_{y} \geq
\frac{m\omega \theta}{2}.
\end{equation}
Eq.(32) shows that for the coherent state (22), the first two
uncertainty relations of (34) are saturated, and the last two
relations of (34) are not (because, as mentioned in Section 2, it
is reasonable to think that $1 \gg \theta$). It is very
interesting that these relations (in eq.(32)) coincides with the
result in ordinary quantum mechanics. In fact, from eqs. (31) one
can see that the noncommutative parameter $\theta$ does not appear
in the expressions of the variances of the operators $\hat{x}$,
$\hat{y}$, $\hat{p}_{x}$ and $\hat{p}_{y}$ in the coherent state
$|\alpha,\beta>$ at all. Here we would like also to point out that
the result in eq.(32) is different from \cite{s17} in which it is
asserted that for a given normalized state, at most one of the
uncertainty relations of NCQM can be saturated, however, in our
present case, there are two of the uncertainty relations are
saturated. Of course, in \cite{s17} only coordinate-coordinate
non-commutativity is discussed, but in the present work, both
coordinate-coordinate and momentum-momentum non-commutativity are
considered.

\section{Summary and discussion}
In this work we consider the structure of Fock space for identical
bosons in a noncommutative space. From the NCQM algebraic
relations (in eq.(1)) for the phase-space variables with
simultaneously coordinate-coordinate and momentum-momentum
non-commutativity, consistent algebraic relations (4) and (6) for
the non-perturbation creation and annihilation operators can be
obtained, which are also regarded as some kind of deformation of
usual bosonic algebraic relations with the deformation parameter
$\theta$.

\par
The number operators $\hat{N}_{a}$ and $\hat{N}_{b}$ are defined
as usual (in eq.(7)). Construction of the noncommutative Fock
space is similar to the case of commutative space, however, unlike
the case of ordinary Fock space, the states $|n,m>$ are
eigenvectors of neither $\hat{N}_{a}$ nor $\hat{N}_{b}$. Using the
Schmidt's orthogonalization method and up to the first power of
the deformation parameter $\theta$, we construct two sets of
orthonormal and complete state-vectors which are simultaneously
eigenvectors of the Hermitian operator
$\hat{N}_{a}+\hat{N}_{b}-\theta \hat{L}$ and $\hat{N}_{a}$ or
$\hat{N}_{b}$ and can be used as bases for the noncommutative Fock
space. Besides this, we also find out other set of bases, such as
$|n,m>_{a,b}$ in eq.(20).

\par
In the noncommutative Fock space, we also construct the
two-dimensional canonical coherent states $|\alpha,\beta>$. It is
worth noting that for $|\alpha,\beta>$ the uncertainty relations
of the noncommutative coordinate and momentum variables take the
exactly same expressions as ones in ordinary quantum mechanics (in
eq.(32)), which imply that the conclusion in \cite{s17} should be
modified in both coordinate-coordinate and momentum-momentum
noncommutative case.

\par
In this work we only investigate some essentially new features of
the noncommutative Fock space. Since the notion of the Fock space
plays an important role in quantum mechanics, it is deserved to
study the noncommutative Fock space further.

\end{document}